\documentclass[showpacs,aps,prb,twocolumn,superscriptaddress,10pt]{revtex4-2}
\usepackage{graphicx} % Include figure files
\usepackage{bm}% bold math
\usepackage{color}

\usepackage{amsmath}
\usepackage{amssymb}
\usepackage{enumerate}
\usepackage{xspace}
\usepackage{mathrsfs}
\usepackage{mathptmx}
\usepackage[utf8]{inputenc}
\usepackage{longtable}
\usepackage{orcidlink}

\newcommand{\smJij}{\ensuremath{J_{ij}}\xspace}

\newcommand{\sms}{\ensuremath{\mathbf{S}}\xspace}

\newcommand{\MnAu}{\ensuremath{\mathrm{Mn}_2\mathrm{Au}}\xspace}
\newcommand{\NdFeB}{\ensuremath{\mathrm{Nd}_2\mathrm{Fe}_{14}}B\xspace}
\newcommand{\Neel}{N\'eel\xspace}
\newcommand{\uvecx}{\mathbf{\hat{x}}}
\newcommand{\uvecy}{\mathbf{\hat{y}}}
\newcommand{\uvecz}{\mathbf{\hat{z}}}
\newcommand{\Del}{\boldsymbol{\nabla}}
\newcommand{\heff}{\mathbf{B}_{\mathrm{eff}}}

\hypersetup{
    breaklinks=true,
    colorlinks=true,
    urlcolor=blue,
    linkcolor=blue,
    citecolor=blue}

\begin{document}

\title{Generalised form of the magnetic anisotropy field in micromagnetic and atomistic spin models}
\author{Jack~B.~Collings~\orcidlink{0000-0002-9354-2903}}
\email{jbc525@york.ac.uk}
\affiliation{Department of Physics, University of York, York, YO10 5DD, UK}
\author{Ricardo~Rama-Eiroa~\orcidlink{0000-0002-7744-6487}}
\affiliation{Donostia International Physics Center, 20018 San Sebasti\'{a}n, Spain}
\affiliation{Polymers and Advanced Materials Department: Physics, Chemistry, and Technology, University of the Basque Country, UPV/EHU, 20018 San Sebasti\'{a}n, Spain}
\affiliation{EHU Quantum Center, University of the Basque Country, UPV/EHU, 48940 Leioa, Spain\looseness=-1}
\author{Rub\'en M.~Otxoa~\orcidlink{0000-0003-1534-4159}}
\affiliation{Hitachi Cambridge Laboratory, J. J. Thomson Avenue, Cambridge CB3 0HE, United Kingdom}
\affiliation{Donostia International Physics Center, 20018 San Sebasti\'{a}n, Spain}
\author{Richard~F.~L.~Evans~\orcidlink{0000-0002-2378-8203}}
\affiliation{Department of Physics, University of York, York, YO10 5DD, UK}
\author{Roy~W.~Chantrell~\orcidlink{0000-0001-5410-5615}}
\email{roy.chantrell@york.ac.uk}
\affiliation{Department of Physics, University of York, York, YO10 5DD, UK}

\begin{abstract}
We present a general approach to the derivation of the effective anisotropy field which determines the dynamical behaviour of magnetic spins according to the Landau-Lifshitz-Gilbert equation. The approach is based on the gradient in spherical polar coordinates with the final results being expressed in Cartesian coordinates as usually applied in atomistic and micromagnetic model calculations. The approach is generally valid for all orders of anisotropies including higher order combinations of azimuthal and rotational anisotropies often found in functional magnetic materials such as permanent magnets and an emerging class of antiferromagnetic materials with applications in spintronics. Anisotropies are represented in terms of spherical harmonics which have the important property of rational temperature scaling. Effective field vectors are given for anisotropies up to sixth order, presenting a unified framework for implementing higher order magnetic anisotropies in numerical simulations. 
\end{abstract}
\maketitle
%\pacs{75.10.Hk,75.20.-g,75.50.Ss,75.60.Jk,75.78.Jp}

\section{Introduction}

Magnetic materials are central to developments which underpin today's information based society.
Applications range from magnetic information storage, still the dominant form of cloud storage, to the exciting possibilities associated with spintronics~\cite{Dieny2020}.
The transition to hybrid and all-electric vehicles is driving renewed interest in the development of high-performance permanent magnets, with a particular aim of reducing the quantity of strategic rare-earth materials.
Finally, the development of 'Neuromorphic spintronics' raises the exciting possibility of the use of magnetic textures as neurons in brain-inspired computing and the  use of magnetic tunnel junctions to simulate both synapses and neurons~\cite{Grollier2020}.
Neuromorphic spintronics can combine computation and memory at a very local level, reducing power requirements, and provides various physical mechanisms as a source of computational power.

A feature of many of these applications is the use of magnetic materials with higher-order magnetic anisotropies due to their composite structure or the intrinsic properties of the material.
An archetypal example is the alloy \NdFeB which revolutionised the permanent magnet industry from the mid 1980s.
Recently the drive for improved motors has led to the use of similar alloys in complex structures aimed at increasing the magnet energy product while reducing the rare-earth content.
More recently, with the burgeoning interest in antiferromagnetic spintronics, it has been shown that using the antiferromagnet as the active element instead of a ferromagnet the device can significantly outperform conventional devices~\cite{MacDonald2011,Gomonay2014,Jungwirth2016,BaltzRevModPhys2018,Jungwirth2018,Jungfleisch2018,Fukami2020}.
One of the most promising materials for these devices is \MnAu due to its high \Neel temperature, moderate anisotropy and layered multi-sublattice spin structure.
It has been predicted that in certain antiferromagnets (AFMs) a current induced spin-orbit-torque is able to switch the sublattice magnetic orientation~\cite{MacDonald2011,Gomonay2014}.
Recently this has been demonstrated experimentally for CuMnAs~\cite{Wadley2016,Olejnk2017,Olejnk2018,MatallaWagnerPRA2019} and \MnAu~\cite{MeinertPRApp2018}.
A common feature of both \NdFeB and \MnAu is the presence of higher order anisotropies, including fourth order rotational terms.
In many materials, including \MnAu~\cite{Shick2010}, the anisotropy arises from anisotropic exchange and has a 2-site form~\cite{EvansPRBR2020}.
The physical understanding of such complex materials and in particular the temperature dependence of properties requires atomistic spin models which are capable of modelling higher order anisotropies and  anisotropic exchange.

Atomistic spin models~\cite{vampire} are based on classical spins interacting via the use of a tensor form of the exchange interaction which introduces e.g., the Heisenberg interaction, exchange anisotropy, and the Dzyaloshinskii–Moriya interaction.
Static properties such as the temperature dependence of the magnetisation can be investigated using standard Metropolis Monte-Carlo methods~\cite{Metropolis1953}.
Calculations of the magnetic anisotropy and its temperature dependence are carried out using the constrained Monte-Carlo (CMC) method of Asselin et.al.~\cite{AsselinPRB2010}.
Spin dynamic calculations are also possible using integration of the stochastic Landau-Lifshitz-Gilbert (LLG) equation.
Characterisation of the temperature dependence of the anisotropy is extremely important for applications which involve elevated temperatures, such as permanent magnets in motor applications and heat assisted magnetic recording (HAMR)~\cite{Rottmayer2006} where the recording medium is heated to beyond the magnetic ordering (Curie) temperature to achieve writing.
It is important to note that the temperature variation of the macroscopic anisotropy $K(T)$ is a thermodynamic free energy: the intrinsic anisotropy at the atomic level has a quantum mechanical quantity which varies slowly with temperature. To characterise $K(T)$ the Akulov-Zener-Callen-Callen law~\cite{AkulovZP1936,ZenerPRB1954,Callen1966} describing the magnetization scaling of the temperature dependent magnetic anisotropy is often used:
\begin{equation}
\frac{ K \left( T \right)}{K \left(T = 0 \right)} = \left( \frac{M \left( T \right) }{M \left( T = 0 \right) } \right)^{ p \left( p + 1 \right) / 2 },
\label{eq:k4rT}
\end{equation}
where $M$ is the bulk magnetisation, and $p$ is the power order of the anisotropy.
However it is important to note that this scaling applies only when the anisotropy is represented by orthogonal functions~\cite{Callen1966}. 

Experimentally magnetic anisotropy is measured indirectly, considering the restoring torque of a magnetic sample where the magnetic moment is oriented away from a known crystallographic axis.
For a uniaxial system these curves are then fitted assuming a form of the anisotropy energy $E_k$ given by
%----------------------------
\begin{equation}
E_k = K_1 \sin^2\theta + K_2 \sin^4 \theta + K_3 \sin^6 \theta
\label{eq:exptK}
\end{equation}
%----------------------------
where $K_1, K_2, K_3$ are the first, second, and third anisotropy constants respectively.
Originally the non-orthogonality of these functions was not considered important, but when considering the temperature dependence of the constituent terms, the actual scaling depends on the relative values of the anisotropy constants~\cite{Skomski2008}.
Another common misconception in the literature is confusing the number of the anisotropy constant with the order of the effect. The first term in Eq.~\eqref{eq:exptK} is second order in $\theta$ (or equivalently in magnetisation $m_z$) and is not a first order anisotropy. Similarly the second term is $4^{\mathrm{th}}$ order, and the third term is $6^{\mathrm{th}}$ order.
This distinction is important when considering the scaling of magnetic anisotropy with temperature and so we make a specific point of clarification here. 

Here we present a unified framework for the implementation of the full range of magnetic anisotropies, expressed in terms of spherical harmonics, in atomistic spin models that is self consistent and follows the analytically derived scaling laws for their separable components.
The derived fields work for both the Landau-Lifshitz-Bloch (LLB) and LLG integration schemes and a further technique allows for an effective field which is more computationally efficient, but can only be used for the LLG formalism.
The LLG form of the fields have been fully implemented in the \textsc{vampire} open source software package \cite{vampireURL,vampire}.
It is important to note that both approaches: the power law expression~\ref{eq:exptK} and the use of spherical harmonics presented here are both perfectly valid.
Eq.\ref{eq:exptK}, when fitted to experimental data gives values of the coefficients which fully characterise the anisotropy energy surface and its temperature dependence for a given material.
Skomskii~\cite{Skomski2008} shows that the coefficients of both representations can be simply related.
However, the temperature dependence of the coefficients determined by fitting to Eq.~\ref{eq:exptK} will not have the rational scaling with magnetization of a fitting to spherical harmonics because of the requirement of representation (ref~\cite{Callen1966}) in terms of orthogonal functions.
Where this scaling is important, the representation in terms of spherical harmonics is required.

\section{Theory}

\subsection{General framework of atomistic spin models}
The general formulation of atomistic spin models \cite{vampire} employs a lattice of fixed length atomic spins interacting with a spin Hamiltonian of the form 
%----------------------------
\begin{equation}
\mathscr{H} = - \sum_{i < j} \smJij \sms_i \cdot \sms_j + \mathscr{H}_{\mathrm{A}} - \sum_i \mu_i \mathbf{B} \cdot \sms_i 
\label{eq:hamiltonian}
\end{equation}
%----------------------------
where $\sms_{i,j}$ are unit vectors describing the direction of local moments $i,j$, $J_{ij}$ is the Heisenberg exchange interaction between neighbouring spins, $\mu_i$ is the spin moment of atom $i$ and $\mathbf{B}$ is the externally applied magnetic induction field. The focus of this article is anisotropic contributions, represented by the general contribution to the spin Hamiltonian $\mathscr{H}_{\mathrm{A}}$. We will present detailed expressions for this term in later sections.

\subsection{Uniaxial and rotational anisotropy}

In this formalism for anisotropy, the easy/hard axis vector is taken to lie along the $\mathbf{\hat{z}}$~axis of the unit cell, from which $\theta$ is measured, and the $\mathbf{\hat{x}}$~axis determines the zero for the measurement of $\phi$ in accordance with the standard spherical polar coordinate system.
Ignoring the effects of neighbouring moments, the anisotropic energy of a magnetic moment $\mathbf{m}$ is some function $E_A \left( r, \theta, \phi \right)$ where $r = ||\mathbf{m}||$ commonly written as $\mu$.
Completely ignoring the effects of neighbouring moments may mean that circumstances in which high changes in magnetisation direction occurring over small distances are not well described by the model, although to a degree this may be accounted for by including two ion anisotropy.
The magnetic field is therefore found by taking the negative gradient $- \boldsymbol{\nabla{E_A}}$ w.r.t. $\mathbf{m}$.
In general, any real scalar function on a spherical surface can be written in terms of real spherical-harmonics such that $Y \left( \theta, \phi \right) = \sum_{l, m} k_{l, m} Y_{l, m} \left( \theta, \phi \right)$, with the standard $m$ and $l$ integers where $|m| \leq l$ and $l \geq 0$, and the $k_{l, m}$ are constants.
To conserve time reversal symmetry, invariance of $Y \left( \theta, \phi \right)$ w.r.t. the combined transformations $\theta \rightarrow \pi - \theta$ and $\phi \rightarrow \phi + \pi$ is required, thus $m + l$ must be even.
Spherical harmonics satisfying these requirements can describe triclinic crystal systems.
The symmetries of other crystal systems allow even more simplifications to be made.
For many systems, the $x$-axis can be chosen such that there is symmetry w.r.t. $\phi$, allowing all negative $m$ terms to be omitted.
In addition, if the system is orthorhombic there is order two rotational symmetry, thus only even $m$, and consequently even $l$ are allowed.
Tetragonal systems have order four rotational symmetry only permitting $m$ to be a multiple of four.
Cubic systems, in addition to the tetragonal constraint, fix the ratios between the harmonic terms such that fourth order rotational symmetry about the $x$ and $y$ axes is obeyed.
Hexagonal systems with sixth order rotational symmetry only permit $m$ to be multiples of six.
Table \ref{tab: energies} gives a subset of the allowed terms likely to be useful to such systems.
The constants have been chosen such that the highest order trigonometric terms are pre-multiplied by $-1$.
Since only energy differences exhibit physical properties, any constant energies have also been removed.

To include the radial dependence, the most general form involves multiplying each of the spherical harmonic terms by some function of $r$.
Understanding the radial dependence is nontrivial due to the difficultly in observing the spin-length in magnetic materials.
Here the dependence can be expressed generally as a Maclaurin expansion, where since symmetry of $E_A$ is required for $r \rightarrow -r$, only even power terms in $r$ are non-zero.
Multiplying the $l = 0$ harmonic by such a polynomial yields a form similar to that of the Landau Hamiltonian.
The form is restricted to a Maclaurin expansion, since the presence of diverging field energies would be un-physical suggesting a Laurent series is inappropriate.
However, it is worth noting that a Maclaurin expansion does not have to be used, as any orthogonal expansion would work. 
It may therefore be useful to consider that restriction to a Maclaurin formalism is not required and in fact any function of $|r|$ is allowed as long as it is piecewise smooth.
Considering the general case of a Maclaurin expansion, the final form can be written as
\begin{equation}\label{eq: general-anisotropic-hamiltonian}
    E_A(r, \theta, \phi) = \sum_{n, l, |m|} k_{n,l,m}r^nY_{l, m}(\theta, \phi)
\end{equation}
where the integers $n, l \geq 0$, $n, (l + m)$ are even, $|m| \leq l$ and $k_{n, l, m}$ are constants.

It is important to note that the only ways to obtain the necessary constants are through the use of either \textit{ab initio} or experimental methods.

\section{Effective field representation}
The effective anisotropy field can be defined as the field which acts on the magnetisation to give the torque
\begin{equation}
    \boldsymbol{\Gamma} = \mathbf{M} \times \heff .
    \label{eq:torque1}
\end{equation}
However, according to Brown~\cite{Brown1962}, the field is indeterminate by a factor $\lambda \mathbf{M}$ where $\lambda$ is any scalar. Further, the effective field cannot be considered an external field since the anisotropy energy cannot be expressed in terms of $\heff$ from Eq.~\ref{eq:torque1} as
\begin{equation}
    E_{A}= -\mathbf{M} \cdot \heff .
    \label{eq:energy1}
\end{equation}
Fujiwara and Zhao~\cite{Fujiwara2000} tackled this problem directly by seeking an expression for $\heff $ which simultaneously satisfied both Eq.~\ref{eq:torque1} and~\ref{eq:energy1}. They solved the problem by  constructing the Stoner-Wohlfarth~\cite{Stoner1948} astroid and determining a simultaneous solution
of Eq.~\ref{eq:torque1} and~\ref{eq:energy1} as the intersection of 2 vectors constructed from the astroid. However, it is straightforward for purely rotational anisotropy to analytically solve Eq.~\ref{eq:torque1} and~\ref{eq:energy1}. For an $n$-fold rotational anisotropy, $E= K\cos(n\phi)$  and $\Gamma=-nK\sin(n\phi)$, and direct solution gives the following expressions for the components of the anisotropy field, $B_{x}$ and $B_{y}$ as
\begin{equation}
    \begin{split}
        B_{x} &= \frac{1}{2} B_K \left[ \cos\left( n \phi \right) \cos\phi + n \sin\left( n \phi \right) \sin\phi \right] , \\
        B_{y} &= \frac{1}{2} B_K \left[ \cos\left( n \phi \right) \sin\phi - n \sin\left( n \phi \right) \cos\phi \right],
        \label{eq:field_compts}
    \end{split}
\end{equation}
where $B_K = 2K / M_s$. 
In terms of Cartesian coordinates the 4th order anisotropy field components become
\begin{equation}
    \begin{split}
        B_{x}&=\frac{B_K}{2}m_x [16m_y^2(m_x^2-m_y^2) +8m_x^4-8m_x^2+1] \\
        B_{y}&=\frac{B_K}{2}m_y [-16m_x^2(m_x^2-m_y^2) +8m_x^4-8m_x^2+1].
        \label{eq:cartesian}
    \end{split}
\end{equation}
This is relatively cumbersome for such a simple form of anisotropy. Also, observation of Eq.~\ref{eq:field_compts} shows that each effective field component contains a term in the energy $\cos(n\phi)$ and the torque $n\sin(n\phi)$: the dot or cross product simply selects respectively the former or the latter. Any such derivation will produce a similarly non-physical field.
The next section will outline a technique which goes further to derive fields which work generally for spin vectors not just in a plane, but any direction on the three dimensional sphere.
Such fields derived would be appropriate for use with the LLB equation, and a technique to obtain simplified effective fields, which are less computationally demanding and appropriate for use with LLG integration, will also be discussed.

\section{General approach to rotational anisotropies}
We start with the gradient in spherical polar coordinates w.r.t. $\mathbf{m}$, equivalent to $ \partial / \partial{\mathbf{m}} $ which is given by
\begin{equation}\label{eq: spherical-polar-grad}
\begin{aligned}
    \Del &= \uvecx \left(\sin{\theta}\cos{\phi}\frac{\partial}{\partial{r}} + \frac{\cos{\theta}\cos{\phi}}{r}\frac{\partial}{\partial{\theta}} - \frac{\sin{\phi}}{r\sin{\theta}}\frac{\partial}{\partial{\phi}} \right) \\
    & + \uvecy \left( \sin{\theta}\sin{\phi}\frac{\partial}{\partial{r}} + \frac{\cos{\theta}\sin{\phi}}{r}\frac{\partial}{\partial{\theta}} + \frac{\cos{\phi}}{r\sin{\theta}}\frac{\partial}{\partial{\phi}} \right) \\ 
    & + \uvecz \left(\cos{\theta}\frac{\partial}{\partial{r}}- \frac{\sin{\theta}}{r}\frac{\partial}{\partial{\theta}} \right)
\end{aligned}
\end{equation}
which can be used to find $- \mathbf{\Del} E_A \left(r, \theta, \phi \right)$ for evolution of systems with longitudinal as well as rotational degrees of freedom such described by the LLB equation of motion~\cite{Garanin1997} and its stochastic form~\cite{Chubykalo2006}.
Similarly, models including longitudinal spin fluctuations of the magnetic moment describe changes of spin magnitude using a Landau Hamiltonian with terms up to $m^6$~\cite{Ma2012,Ellis2019}.
In such cases the radial derivatives give the appropriate field.

It is worth noting that terms in Eq.~\eqref{eq: general-anisotropic-hamiltonian} with $n = 0$ give purely anisotropic terms, and terms with $l = 0$ give purely radial terms.
If Landau terms dominate over coupled radial and angular terms, i.e., $|k_{n > 0, 0, 0}| \gg |k_{n > 0, l > 0, m}|$, then perturbations in $r$ have little angular dependence.
Similarly if purely anisotropic terms dominate over coupled angular-radial terms, perturbations in the angular directions have little dependence on $r$.
If both cases are valid, then the dynamics are described by purely anisotropic and radial terms.
An even further simplification is possible if a moment has its radius restricted to a length $r = \mu$ so the expression reduces to an expansion in spherical harmonics with constants $k_{l, m}$, and the LLG equation of motion appropriately describes the system dynamics with
\begin{equation}\label{eq: gilbert-anisotropy-energy}
    E_A \left(r, \theta, \phi \right) = E_A \left( \theta, \phi \right) = \sum_{l, m} k_{l, m}Y_{l, m} \left( \theta, \phi \right).
\end{equation}
For anisotropy energies involving these purely anisotropic terms, there is no radial dependence and the magnetic moment can be simplified with the restriction that it takes a position on a sphere of radius $\mu$ so that Eq.~\ref{eq: spherical-polar-grad} reduces to
\begin{equation}\label{eq: spin-sphere-grad}
\begin{aligned}
    \Del &= \frac{1}{\mu} \left[ \uvecx\left( \cos{\phi}\cos\theta\frac{\partial}{\partial{\theta}} - \frac{\sin{\phi}}{\sin{\theta}}\frac{\partial}{\partial{\phi}} \right) \right. \\
    & \left. + \uvecy\left( \sin{\phi}\cos{\theta}\frac{\partial}{\partial{\theta}} + \frac{\cos{\phi}}{\sin{\theta}}\frac{\partial}{\partial{\phi}} \right) - \uvecz\sin{\theta}\frac{\partial}{\partial{\theta}} \right],
\end{aligned}
\end{equation}
note that this $\mu$ length simplification must not be used for any terms with radial dependence.
From this gradient, the torque can be calculated using $- \mathbf{\hat{m}} \times \mathbf{\nabla} E_{l, m}$, where $E_{l, m}$ represents a contribution due to a purely anisotropic term with the corresponding $l$ and $m$ from $E_A(\theta, \phi)$.
This represents a first principles approach to the determination of effective anisotropy fields for spin dynamics and micromagnetic models.
It is also possible to yield the same results for the fields by representing the anisotropy energies explicitly in terms of direction cosine components, and taking the gradient in the Cartesian basis.
In the following we present details of the calculation for second order anisotropy including the method for the determination of the simplest field representation bearing in mind the indeterminacy of the anisotropy field mentioned by Brown~\cite{Brown1962}.
Following this we derive the expression for fourth order anisotropy including some of the rotational terms.
The results pertaining to other common anisotropies, including the Cartesian representation most relevant for atomistic spin models, are given in Tables \ref{tab: effective_fields}, \ref{tab: torques} and \ref{tab: full_fields}.

\subsection{Second Order}
The second order uniaxial term creates an energetic incentive for the magnetisation to align with the $z$-axis if the second order uniaxial anisotropy constant, $k_{2,0}$, is positive. The form used is given by
\begin{equation}\label{eq: uniaxial-second-order}
    E_{2,0} = -k_{2,0}\cos^{2}{\theta}
\end{equation}
with coefficients and constant offsets removed for simplicity, since it is only gradients in energy which give rise to fields. A similar convention is used for the higher order terms later considered.
A calculation of $-\mathbf{\nabla}E_{2,0}$ using Eq.\ref{eq: spin-sphere-grad} yields
\begin{equation}
    \mathbf{B}_{2,0} = - \frac{2k_{2,0}m_z}{\mu} \left[
    \begin{bmatrix}
        0 \\
        0 \\
        -1
    \end{bmatrix} + m_z
    \begin{bmatrix}
        m_x \\
        m_y \\
        m_z
    \end{bmatrix} \right]
\end{equation}
with $m_x$, $m_y$ and $m_z$ denoting the $\uvecx, \uvecy, \uvecz$ components of $\hat{\mathbf{m}}$ respectively.
The torque is then found from $\mu \hat{\mathbf{m}} \times \mathbf{B}_{2,0}$ to be 
\begin{equation}
    \boldsymbol{\tau}_{2,0} = 2k_{2,0}m_z
    \begin{bmatrix}
    m_y \\
    -m_x \\
    0
    \end{bmatrix}
\end{equation}
so the effective field can be written as
\begin{equation}
    \mathbf{B}_{2,0 \text{ eff}} = \frac{2k_{2,0}}{\mu} m_z
    \begin{bmatrix}
    0 \\
    0 \\
    1
    \end{bmatrix}
\end{equation}
since the part of $\mathbf{B}_{2,0}$ which is proportional to $\hat{\mathbf{m}}$ makes no contribution to the torque. This is the usual expression found by taking the derivative of the energy in Cartesian coordinates. However, this result is coincidental and not generally applicable to higher order magnetic anisotropies. Notice that $\mu \hat{\mathbf{m}} \times \mathbf{B}_{2,0 \text{ eff}}$ yields the same torque as $\mu \hat{\mathbf{m}} \times \mathbf{B}_{2,0}$, and any effective field would obey the equation
\begin{equation}
    \mu
    \begin{bmatrix}
        0 & -m_z & m_y \\
        m_z & 0 & -m_x \\
        -m_y & m_x & 0
    \end{bmatrix}
    \begin{bmatrix}
    B_{2x \text{ eff}} \\
    B_{2y \text{ eff}} \\
    B_{2z \text{ eff}}
    \end{bmatrix}
    = 2k_{2,0}m_z
    \begin{bmatrix}
        m_y \\
        -m_x \\
        0
    \end{bmatrix},
\end{equation}
which represents the simultaneous equations yielded from cross product $\mu \hat{\mathbf{m}} \times \mathbf{B}_{2,0 \text{ eff}}$. Noticing that since the determinant of the matrix as well as all Cramer determinants are zero and that the matrix has non-zero co-factors suggests an infinite number of solutions lying on a line.
From the simultaneous equations, the line equation can then be written as
\begin{equation}
     B_{2x \text{ eff}} = \frac{m_x}{m_y}B_{2y \text{ eff}} = -\frac{2k_{2,0} m_x}{\mu} + \frac{m_x}{m_z}B_{2z \text{ eff}}
\end{equation}
and setting $B_{2x \text{ eff}} = B_{2y \text{ eff}} = 0$ yields the given effective field.
The effective field is in the same plane as that defined by $\mathbf{m}$ and $\mathbf{B}_{2,0}$, however its direction and magnitude are changed from $\mathbf{B}_{2,0}$ so as to remove a component for more efficient computation. This means that there is a radial component in $\mathbf{B}_{2,0 \text{ eff}}$, whereas in the true field $\mathbf{B}_{2,0}$ there is not. Thus such effective fields should only be used for computation of LLG dynamics, and not for the LLB where the true fields should be used. It is possible to use the effective field for LLB dynamics, but only when calculating the cross-product term; the dot product term should be zero.

The other second order term is rotational and described by
\begin{equation}
    E_{2,2} = -k_{2,2}\sin^{2}{\theta}\cos{2\phi}
\end{equation}
where $k_{2,2}$ is a constant.
This contribution occurs in lower symmetry systems i.e. where the rotational symmetry in the plane perpendicular to the uniaxial anisotropy axis is of order $2$ e.g. orthorhombic, monoclinic, and triclinic systems.
Using Eq.\ref{eq: spin-sphere-grad} this yields a field of
\begin{equation}
    \mathbf{B}_{2,2} = \frac{2k_{2,2}}{\mu} \left[
    \begin{bmatrix}
        m_x \\
        -m_y \\
        0
    \end{bmatrix}
    - (m_x^2 - m_y^2)
    \begin{bmatrix}
        m_x \\
        m_y \\
        m_z
    \end{bmatrix}
    \right]
\end{equation}
which allows for an effective field of the form
\begin{equation}
    \mathbf{B}_{2, 2 \text{ eff}} = \frac{2k_{2, 2}}{\mu}
    \begin{bmatrix}
    m_x \\
    -m_y \\
    0
    \end{bmatrix}.
\end{equation}

\subsection{Fourth Order}
We now consider higher order magnetic anisotropies where the Cartesian derivative yields incorrect expressions for the effective magnetic fields, particularly in the case of rotational anisotropies. Experimentally the 4th-order magnetic anisotropy is usually expressed in the form
\begin{equation}
E_{4,0} = K_2 \sin^4 \theta
\end{equation}
which as discussed earlier leads to an incorrect scaling as the common scaling laws are derived for orthogonal functions. In the permanent magnet community the anisotropy is often described by a Legendre polynomial of appropriate order in Cartesian coordinates
\begin{equation}
E_{4,0} = -k_{4}\frac{1}{8}\left(35m_z^{4}-30m_z^{2}+3\right)
\end{equation}
where $k_4$ is a constant. Here the requirement for orthogonality is satisfied by design through the addition of a second order component. A similar spherical harmonic representation is given by the expansion of 4th order uniaxial components of Eq.~\ref{eq:shk}
\begin{equation}
E_{4,0} =-k_{4}\frac{3}{16} \sqrt{\frac{1}{\pi}}\left(  35\cos^{4}\theta - 30\cos^{2}\theta + 3\right).
\end{equation}
The more complicated functional form of the Legendre polynomials and spherical harmonics has limited their wider adoption in the community, especially due to the change of definition of the anisotropy constant $K_l$ to a constant of order $k_{2l}$ and the need to fit to more complicated expressions when measuring torque curves~\cite{Skomski2008}. To balance the requirements for simplicity and orthogonality we have adopted a minimally orthogonal function for higher order anisotropies, where the leading term of the relevant order is normalised to 1 and irrelevant constants are also removed. For 4th order anisotropy with non-rotational terms that leads to an expression of the form 
\begin{equation}
    E_{4,0} = -k_{4, 0}\left( \cos^{4}\theta - \frac{6}{7}\cos^{2}\theta \right)
\end{equation}
with $k_{4, 0}$ denoting the fourth order anisotropy constant.
Using a similar technique to the second order derivation, the field and effective field are then given respectively as
\begin{equation}
    \mathbf{B}_{4, 0}
    = - \frac{2k_{4, 0} m_z}{\mu} \left(2m_z^2- \frac{6}{7} \right) \left[
    m_z 
    \begin{bmatrix}
    m_x \\
    m_y \\
    m_z
    \end{bmatrix} - 
    \begin{bmatrix}
    0 \\
    0 \\
    1
    \end{bmatrix} \right],
\end{equation}
and since $\hat{\mathbf{m}} \times \hat{\mathbf{m}} = 0$
\begin{equation}
    \mathbf{B}_{4, 0 \text{ eff}} = \frac{2k_{4, 0} m_z}{\mu} \left(2m_z^2 - \frac{6}{7} \right)
    \begin{bmatrix}
    0\\
    0\\
    1
    \end{bmatrix}
\end{equation}
to yield the same effective torque. Next considering the rotational terms, one with second order rotational symmetry and the other with fourth order rotational symmetry in the $xy$ plane, the second order rotational term is given by
\begin{equation}
    E_{4, 2} = -k_{4, 2}\sin^2{\theta}\left( \cos^2{\theta} - \frac{1}{7} \right)\cos{2\phi}
\end{equation}
where $k_{4, 2}$ represents the fourth order $\theta$ second order $\phi$ anisotropy constant.
This yields a field and an effective field respectively given by
\begin{equation}
\begin{aligned}
    \mathbf{B}_{4, 2} &= \frac{4k_{4, 2}}{\mu} \left[
    \left( m_x^4 - m_y^4 - \frac{3}{7}(m_x^2 - m_y^2) \right)
    \begin{bmatrix}
        m_x \\
        m_y \\
        m_z
    \end{bmatrix} \right. \\
    & \left. + 
    \begin{bmatrix}
    - m_x^3 + \frac{3}{7} m_x \\
    m_y^3 - \frac{3}{7} m_y \\
    0
    \end{bmatrix} \right],
\end{aligned}
\end{equation}
\begin{equation}
    \mathbf{B}_{4, 2 \text{ eff}} = \frac{4k_{4, 2}}{\mu}
    \begin{bmatrix}
        -m^{3}_x + \frac{3}{7}m_x \\
        m^{3}_y - \frac{3}{7}m_y \\
        0
    \end{bmatrix}.
\end{equation}

The fourth order rotational term, which is applicable to $\text{Mn}_2\text{Au}~$\cite{Shick2010}, is given by
\begin{equation}
    E_{4, 4} = -k_{4, 4}\sin^{4}\theta\cos{4\phi}
\end{equation}
where $k_{4, 4}$ represents the fourth order $\theta$ fourth order $\phi$ anisotropy constant. The field and effective field are respectively given as
\begin{equation}
\begin{aligned}
    \mathbf{B}_{4, 4} &= - \frac{4k_{4, 4}}{\mu} \left[
    ( m_x^4 + m_y^4 - 6 m_x^2 m_y^2)
    \begin{bmatrix}
        m_x \\
        m_y \\
        m_z
    \end{bmatrix} \right. \\
    & \left. + 
    \begin{bmatrix}
        m_x(3m_y^2 - m_x^2) \\
        m_y(3m_x^2 - m_y^2)\\
        0
    \end{bmatrix} \right],
\end{aligned}
\end{equation}
\begin{equation}
    \mathbf{B}_{4, 4 \text{ eff}} = \frac{4k_{4, 4}}{\mu}
    \begin{bmatrix}
        m_x (m_x^2 - 3m_y^2)\\
        m_y (m_y^2 - 3m_x^2)\\
        0
    \end{bmatrix}.
\end{equation}
For additional use, the expressions up to 6th order are included in the Appendices in tabulated form. The importance of our approach is demonstrated by the fact that obtaining expressions for the effective magnetic field by calculating the derivative of the energy in Cartesian form yields an incorrect torque and thus incorrect equilibrium and dynamic properties of the system, unless the energy is expressed using direction cosine form before taking the derivatives. Therefore care must be taken when implementing higher order magnetic anisotropies in numerical codes~\cite{donahue1999interagency,Vansteenkiste2014,Skubic2008,spirit,Bisotti2018}.

\section{Conclusions}
Atomistic and micromagnetic models are increasingly used for simulations of the properties of magnetic materials with applications in many fields.
Particularly  challenging is the representation of anisotropy fields for computational dynamic calculation in materials with higher order anisotropies.
Into this consideration comes the important scaling law of magnetic anisotropy which requires representation in terms of orthogonal functions~\cite{Callen1966}.
We have developed a general approach based on the spherical polar representation of the gradient which, coupled with solution of the simultaneous equations yielded from the cross product of the magnetisation and an effective field, leads to  the minimal representation of the effective field for applications in spin dynamics, both in the atomistic and micromagnetic discretization schemes.
The derivation of the Cartesian components of the effective field (most useful for atomistic and micromagnetic simulations) is given for two specific examples, with an extensive collection of results being given in the appendix.
All the examples described here have been included in the VAMPIRE~\cite{vampireURL} code.

\section*{acknowledgments}
The authors would like to thank Andrew Naden for helpful discussions with the formulation of the spherical harmonic form of magnetic anisotropy.
JBC gratefully acknowledges the studentship provided by Seagate Technology.
\section*{author contributions}
JBC and RR-E contributed equally to this work.

\bibliographystyle{apsrev4-1}
\bibliography{library}

\clearpage

\appendix

\section{Tables of Anisotropy Energies, Fields and Torques}

\begin{table}[h!]
\centering
    \begin{tabular}{c c}
        \hline \hline
        Order & Energy \\
        \hline
        2,0 & $-k_{2,0}\cos^2{\theta}$ \\
        2,2 & $-k_{2,2}\sin^2{\theta}\cos{2\phi}$ \\
        4,0 & $-k_{4,0}\left( \cos^4{\theta} - \frac{6}{7}\cos^2{\theta} \right)$ \\
        4,2 & $-k_{4,2}\sin^2{\theta}\left( \cos^2{\theta} - \frac{1}{7} \right)\cos{2\phi}$ \\
        4,4 & $-k_{4,4}\sin^4{\theta}\cos{4\phi}$ \\
        6,0 & $-k_{6,0}\left( \cos^6{\theta} - \frac{15}{11} \cos^4{\theta} + \frac{5}{11} \cos^2{\theta} \right)$ \\
        6,2 & $-k_{6,2} \sin^2{\theta} \left( \cos^4{\theta} - \frac{9}{11}\cos^2{\theta} + \frac{1}{33} \right)\cos{2\phi} $ \\
        6,4 & $-k_{6,4} \sin^4{\theta} \left( \cos^2{\theta} - \frac{1}{11} \right)\cos{4\phi}$ \\
        6,6 & $-k_{6,6} \sin^6{\theta}\cos{6\phi}$ \\
        \hline \hline
    \end{tabular}
\caption{\small Expressions for anisotropy energies given for order $\theta$, order $\phi$.}
\label{tab: energies}
\end{table}

\begin{table}[h!]
\centering
    \begin{tabular}{c c}
        \hline \hline
        Order & Effective Field \\
        \hline
        2,0 & $(2k_{2,0} m_z / \mu)
            \begin{bmatrix}
            0 \\
            0 \\
            1
            \end{bmatrix}$ \\
        2,2 & $(2k_{2, 2} / \mu)
            \begin{bmatrix}
            m_x \\
            -m_y \\
            0
            \end{bmatrix}$ \\
        4,0 & $(2k_{4, 0} m_z / \mu) \left(2m_z^2 - \frac{6}{7} \right)
            \begin{bmatrix}
            0 \\
            0 \\
            1
            \end{bmatrix}$ \\
        4,2 & $(4k_{4, 2} / \mu)
            \begin{bmatrix}
                -m^{3}_x + \frac{3}{7}m_x \\
                m^{3}_y - \frac{3}{7}m_y \\
                0
            \end{bmatrix}$ \\
        4,4 & $(4k_{4, 4} / \mu)
            \begin{bmatrix}
            m_x (m_x^2 - 3m_y^2) \\
            m_y (m_y^2 - 3m_x^2) \\
            0
            \end{bmatrix}$ \\
        6,0 & $(2k_{6, 0} m_z / \mu) \left(3m_z^4 - \frac{30}{11}m_z^2 + \frac{5}{11} \right)
            \begin{bmatrix}
            0 \\
            0 \\
            1
            \end{bmatrix}$ \\
        6,2 & $(2k_{6, 2} / \mu)
            \begin{bmatrix}
            m_y(m^{2}_x + m^{2}_y)(m^{2}_x - 3m^{2}_y) - \frac{32}{11} m^{3}_x + \frac{16}{33}m_x \\
            m_x(m^{2}_x + m^{2}_y)(3m^{2}_x - m^{2}_y) + \frac{32}{11} m^{3}_y - \frac{16}{33}m_y \\
            0
            \end{bmatrix}$ \\
        6,4 & $(2k_{6, 4} / \mu)
            \begin{bmatrix}
            m_x((5m^{4}_y + 10m^{2}_x m^{2}_y - 3m^{4}_x) + \frac{20}{11}(m^{2}_x - 3m^{2}_y)) \\
            m_y((5m^{4}_x + 10m^{2}_x m^{2}_y - 3m^{4}_y) + \frac{20}{11}(m^{2}_y - 3m^{2}_x)) \\
            0
            \end{bmatrix}$ \\
        6,6 & $(6k_{6, 6} / \mu)
            \begin{bmatrix}
            -m_x(10m_y^2m_x^2 - m_x^4 - 5m_y^4) \\
            m_y(10m_x^2m_y^2 - m_y^4 - 5m_x^4) \\
            0
            \end{bmatrix}$ \\
        \hline \hline
    \end{tabular}
\caption{\small Effective anisotropy field expressions given for order $\theta$, order $\phi$.}
\label{tab: effective_fields}
\end{table}

\begin{table}
\centering
    \begin{tabular}{c c}
        \hline \hline
        Order & Torque \\
        \hline
        2,0 & $2k_{2,0} m_z
            \begin{bmatrix}
            m_y \\
            -m_x \\
            0
            \end{bmatrix}$ \\
        2,2 & $2k_{2,2}
            \begin{bmatrix}
            m_y m_z \\
            m_x m_z \\
            -2m_x m_y
            \end{bmatrix}$ \\
        4,0 & $2k_{4,0} m_z \left(2m_z^{2} - \frac{6}{7} \right)
            \begin{bmatrix}
            m_y \\
            -m_x \\
            0
            \end{bmatrix}$ \\
        4,2 & $-4k_{4,2}
            \begin{bmatrix}
            m_y m_z(m^{2}_y - \frac{3}{7}) \\
            m_x m_z(m^{2}_x - \frac{3}{7}) \\
            m_x m_y(-m^{2}_x - m^{2}_y + \frac{6}{7})
            \end{bmatrix}$ \\
        4,4 & $-4k_{4,4}
            \begin{bmatrix}
                m_y m_z (-3m_x^{2} + m_y^{2}) \\
                m_x m_z (3m_y^{2} - m_x^{2}) \\
                4m_x m_y (m_x^{2} - m_y^{2})
            \end{bmatrix}$ \\
        6,0 & $-2k_{6,0} m_z \left(3m_z^{4} - \frac{30}{11}m_z^{2} + \frac{5}{11} \right)
            \begin{bmatrix}
            -m_y \\
            m_x \\
            0
            \end{bmatrix}$ \\
        6,2 & $-2k_{6,2}
            \begin{bmatrix}
                m_y m_z ((m_x^2 + m_y^2)(3m_y^2 - m_x^2) - \frac{32}{11}m_y^2 + \frac{16}{33}) \\
                m_x m_z ((m_x^2 + m_y^2)(3m_x^2 - m_y^2) - \frac{32}{11}m_x^2 + \frac{16}{33}) \\
                -2 m_x m_y((m_x^2 + m_y^2)^2 - \frac{16}{11}(m_x^2 + m_y^2) + \frac{16}{33})
            \end{bmatrix}$ \\
        6,4 & $-2k_{6,4}
        \begin{bmatrix}
            m_y m_z(5 m_x^4 - 10 m_x^2 m_y^2 + 3 m_y^4 + \frac{20}{11}(3 m_x^2 - m_y^2)) \\
            m_x m_z(3 m_x^4 - 10 m_x^2 m_y^2 - 5 m_y^4 - \frac{20}{11}(3 m_y^2 - m_x^2)) \\
            8 m_x m_y(-m_x^4 + m_y^4 - \frac{10}{11}(m_x^2 - m_y^2))
        \end{bmatrix}$ \\
        6,6 & $-6k_{6,6}
        \begin{bmatrix}
            m_y m_z (10 m_x^2 m_y^2 - m_y^4 - 5 m_x^4) \\
            m_x m_z (10 m_x^2 m_y^2 - m_x^4 - 5 m_y^4) \\
            m_x m_y (-20 m_x^2 m_y^2 + 6 m_y^4 + 6 m_x^4)
        \end{bmatrix}$ \\
        \hline \hline
    \end{tabular}
\caption{\small Anisotropy torque expressions given for order $\theta$, order $\phi$.}
\label{tab: torques}
\end{table}

\begin{widetext}

\begin{table}
\centering
    \begin{tabular}{c c}
        \hline \hline
        Order & Field \\
        \hline
        2,0 & $-(2k_{2,0}m_z / \mu)
            \begin{bmatrix}
            m_xm_z \\
            m_ym_z \\
            -(1 - m_z^2)
            \end{bmatrix}$ \\
        2,2 & $(2k_{2,2} / \mu)
            \begin{bmatrix}
            m_x(-m_x^{2} + m_y^{2} + 1) \\
            m_y(-m_x^{2} + m_y^{2} - 1) \\
            m_z(-m_x^{2} + m_y^{2})
            \end{bmatrix}$ \\
        4,0 & $-(2k_{4, 0} m_z / \mu) \left(2m_z^2- \frac{6}{7} \right)
            \begin{bmatrix}
            m_z m_x \\
            m_z m_y \\
            (m_z^2 - 1)
            \end{bmatrix}$ \\
        4,2 & $(4k_{4, 2} / \mu)
            \begin{bmatrix}
            m_x(m^{4}_x - m^{4}_y) - \frac{m_x}{7}(10m^{2}_x - 3m^{2}_y - 7) \\
            m_y(m^{4}_x - m^{4}_y) - \frac{m_y}{7}(3m^{2}_x -10m^{2}_y + 7) \\
            m_z(m^{4}_x - m^{4}_y) - \frac{3}{7}m_z(m_x^{2} - m_y^{2}) \\
            \end{bmatrix}$ \\
        4,4 & $-(4k_{4, 4} / \mu)
            \begin{bmatrix}
            m_x(m_x^4 + m_y^4 - 6 m_x^2m_y^2 + 3m_y^2 - m_x^2) \\
            m_y(m_x^4 + m_y^4 - 6 m_x^2m_y^2 +3m_x^2 - m_y^2)\\
            m_z(m_x^4 + m_y^4 - 6m_x^2m_y^2)
            \end{bmatrix}$ \\
        6,0 & $-(2k_{6, 0} m_z / \mu) \left(3m_z^4 - \frac{30}{11}m_z^2 + \frac{5}{11} \right)
            \begin{bmatrix}
            m_z m_x \\
            m_z m_y \\
            (m_z^2 - 1)
            \end{bmatrix}$ \\
        6,2 & $(2k_{6, 2} / \mu)
            \begin{bmatrix}
            m_x(-3m^{6}_x -3m^{4}_x m^{2}_y + 3m^{2}_x m^{4}_y + 3m^{6}_y + \frac{65}{11}m^{4}_x + 2m^{2}_x m^{2}_y - \frac{43}{11}m^{4}_y - \frac{112}{33}m^{2}_x + \frac{16}{33}m^{2}_y + \frac{16}{33}) \\
            m_y(-3m^{6}_x - 3m^{4}_x m^{2}_y + 3m^{2}_x m^{4}_y + 3m^{6}_y + \frac{43}{11}m^{4}_x - 2m^{2}_x m^{2}_y -\frac{65}{11} m^{4}_y -\frac{16}{33}m^{2}_x + \frac{112}{33}m^{2}_y - \frac{16}{33}) \\
            m_z(m^{2}_x - m^{2}_y)(3(m^{2}_x + m^{2}_y)^2 + \frac{32}{11}(m^{2}_x + m^{2}_y) - \frac{16}{33})
            \end{bmatrix}$ \\
        6,4 & $-(2k_{6, 4} / \mu)
            \begin{bmatrix}
            m_x(3m^{2}_y - m^{2}_x + \frac{53}{11}m^{4}_x - \frac{230}{11}m^{2}_x m^{2}_y - \frac{35}{11}m^{4}_y - 3m^{6}_x - 15m^{4}_x m^{2}_y - 15m^{2}_x m^{4}_y - 3 m^{6}_y) \\
            m_y(3m^{2}_x - m^{2}_y + \frac{53}{11}m^{4}_y - \frac{230}{11}m^{2}_x m^{2}_y - \frac{35}{11}m^{4}_x - 3 m^{6}_y  + 15m^{2}_x m^{4}_y + 15 m^{4}_x m^{2}_y - 3m^{6}_x) \\
            -3m_z((m^{2}_x + m^{2}_y)(m^{4}_x - 6m^{2}_x m^{2}_y + m^{4}_y) + \frac{20}{11}(m_x^{4} + m_y^{4} - 6m_x^{2}m_y^{2}))
            \end{bmatrix}$ \\
        6,6 & $-(6k_{6, 6} / \mu)
            \begin{bmatrix}
            m_x(m_x^6 - 15m_x^4m_y^2 + 15m_x^2m_y^4 - m_y^6 + 10m_x^2m_y^2 - m_x^4 - 5m_y^4) \\
            m_y(m_x^6 - 15m_x^4m_y^2 + 15m_x^2m_y^4 - m_y^6 - 10m_x^2m_y^2 + m_y^4 + 5m_x^4) \\
            m_z(m_x^6 - 15m_x^4m_y^2 + 15m_x^2m_y^4 - m_y^6)
            \end{bmatrix}$ \\
        \hline \hline
    \end{tabular}
\caption{\small Anisotropy field expressions given for order $\theta$, order $\phi$.}
\label{tab: full_fields}
\end{table}

\end{widetext}

%\clearpage
%\bibliography{/Users/rfle500/Documents/Work/Papers/Bibliography/library}
%\bibliography{library,local}

%---------------------------------------------------------------------------%

%\begin{thebibliography}{11}%
%\end{thebibliography}%

\end{document}